\documentclass[preprint,12pt]{elsarticle}

\usepackage{graphicx}
\usepackage{amsmath,amssymb}
\usepackage{booktabs}
\usepackage{array}
\usepackage{multirow}
\usepackage{hyperref}
\usepackage{url}
\usepackage{float}
\usepackage{enumitem}
\usepackage{placeins}
\newcommand{\kms}{km\,s$^{-1}$}
\newcommand{\omeganative}{km\,s$^{-1}$\,kpc$^{-1}$}
\newcommand{\Vobs}{\ensuremath{V_{\mathrm{obs}}}}

\newcommand{\Vbary}{\ensuremath{V_{\mathrm{bary}}}}

\newcommand{\DD}{\mathrm{d}}

\journal{Astronomy and Computing}

\begin{document}
\begin{frontmatter}

\title{A Reproducible Two-Boundary Kinematic Correction for Baryonic Rotation-Curve Reconstruction in an 84-Galaxy SPARC Benchmark}

\author{D.\,C.~Flynn}
\ead{davidflynn@eps-research.com}
\address{EPS Research, Laurel, MD 20707, USA}

\begin{abstract}
We present a reproducible computational validation and failure analysis
of the empirical $\omega$ kinematic correction introduced by Flynn and
Cannaliato (2025).  The algorithm is deliberately minimal: one
coefficient per galaxy is calculated from the innermost and outermost
measured rotation-curve points and applied to the full observed radial
profile before comparison with a baryonic reconstruction assembled from
SPARC gas, disk, and bulge components.  We preserve the predecessor's
frozen 84-galaxy benchmark and publish its exact membership so that the
transformation, not sample re-selection, is the object of validation.
Without fitting the transformation to the baryonic residual, the primary
maximum-disk reconstruction reduces the mean observed--baryonic discrepancy
from 51.82 to 30.15~\kms across the frozen 84-galaxy benchmark.  Bounded
mass-to-light-ratio optimization further reduces the descriptive
sensitivity-fit value to 25.45~\kms, while the simple Keplerian reference
has a mean RMSE of 74.20~\kms in our earlier analysis~\cite{Flynn2025}.  Recalculation
from the 84 per-galaxy records shows a resolved mass-to-light optimization benefit
($\Delta\mathrm{RMSE}>0.05$~\kms) in 53 galaxies and no resolved change
in 31.  Six galaxies do not beat the Keplerian reference; all
six occur at $\Upsilon_{\max}\leq0.111$, whereas their $\omega$ values
are not concentrated at the high end of the sample.  We specify the
complete deterministic workflow, native units, endpoint invariants,
uncertainty propagation, and formula-level regression checks required
to prevent grouping and sign errors.  The complete 84-galaxy panel set,
population-level error distributions, and failure diagnostics are
retained as inspectable outputs.  The result is a reproducible
astronomical data-transformation benchmark rather than a proposed force
law or replacement for dark matter or modified gravity.
\end{abstract}


\end{frontmatter}

\section{Introduction}
\label{sec:intro}

Resolved galaxy rotation curves are a standard test bed for both
astrophysical modelling and astronomical data analysis.  The observed
circular velocity commonly remains above the velocity associated with
the measured baryonic components at large radius
\cite{Rubin1980,vanAlbada1985}.  In the standard $\Lambda$CDM framework
this difference is modelled with dark-matter haloes
\cite{NFW1997,Begeman1991}; MOND instead modifies the low-acceleration
relation between baryonic and observed dynamics
\cite{Milgrom1983,Sanders2002,FamaeySanders2012}.  Large public data sets
such as SPARC \cite{Lelli2016,Lelli2017}, THINGS
\cite{Walter2008,deBlok2008}, and LITTLE THINGS \cite{Oh2015} make the
same problem useful from a computational perspective: a proposed
transformation can be applied to many resolved curves, its numerical
assumptions can be made explicit, and its outputs can be independently
reproduced.

Flynn and Cannaliato \cite{Flynn2025} introduced an empirical
transformation based on a two-boundary coefficient $\omega$.  The
original objective was intentionally narrow: starting from an observed
rotation curve, determine whether a correction derived only from the
inner and outer measured points can reproduce the characteristic
Keplerian-like decline used as a reference in the predecessor study.
The resulting curves were visually compared with the well-known M33
rotation-curve morphology and then extended across a SPARC sample.  The
present paper asks the next, stronger question: when the same
transformation is evaluated against the independently supplied SPARC
baryonic components, what does the numerical reconstruction actually
recover?

The contribution of this paper is therefore computational rather than
cosmological.  We provide: (i) an explicit, dimensionally consistent
statement of the two-boundary estimator; (ii) a reproducible workflow
for SPARC component assembly and bounded mass-to-light-ratio sensitivity
analysis; (iii) full-sample visual diagnostics for the 84-galaxy
benchmark used by the predecessor study; and (iv) numerical invariants
that distinguish the canonical estimator from a previously encountered
mis-parenthesized variant.  This framing is aligned with the role of
astronomical computing as a means to make data transformations,
validation steps, and software assumptions inspectable and repeatable.

\section{Data and benchmark definition}
\label{sec:data}

\subsection{SPARC inputs}

SPARC \cite{Lelli2016} provides resolved rotation curves and baryonic
mass-model components at 3.6~$\mu$m.  The analysis uses radius $R$,
observed circular velocity $V_{\rm obs}$, its reported uncertainty, and
the gas, stellar-disk, and bulge velocity contributions $V_g$, $V_d$,
and $V_b$.

\subsection{Frozen 84-galaxy benchmark}
\label{sec:benchmark}

For direct continuity with \cite{Flynn2025}, this study uses the same
84-galaxy benchmark carried forward from that work.  The predecessor selected
SPARC quality flag $Q=1$, required at least ten radial data points and non-zero
radius, and then retained the published 84-member cohort used for the original
validation.  The exact galaxy
identifiers are enumerated in Tables~\ref{tab:upsilon} and
\ref{tab:upsilon2} and in the deposited analysis products.  The cohort
is held fixed throughout the present study: no galaxy is added, removed,
or reclassified during the baryonic reconstruction.  Treating cohort
membership as an explicit input is part of the computational
specification and ensures that the present experiment tests the
transformation rather than a revised sample-selection rule.

\section{Computational method}
\label{sec:method}

\subsection{Canonical $\omega$ estimator}

Let $(R_1,V_1)$ and $(R_2,V_2)$ denote the innermost and outermost
measured points of a galaxy rotation curve.  The canonical estimator is
\begin{equation}
  \boxed{\omega = \frac{V_2}{R_2}
  - \frac{V_1}{R_1}\left(\frac{R_1}{R_2}\right)^{3/2}}
  \label{eq:omega}
\end{equation}
and the associated Keplerian reference and transformed velocity are
\begin{align}
  V_{\mathrm{Kep}}(R) &= V_1\left(\frac{R_1}{R}\right)^{1/2},
  \label{eq:vkep}\\
  V_{\mathrm{adj}}(R) &= V_{\mathrm{obs}}(R)-R\omega.
  \label{eq:vadj}
\end{align}
Equation~\ref{eq:omega} is the corrected and locked form of Eq.~6 in
\cite{Flynn2025}.  In executable notation it is
\begin{equation}
 \omega = (V_2/R_2) - (V_1/R_1)(R_1/R_2)^{1.5},
\end{equation}
not
$[(V_2/R_2)-(V_1/R_1)](R_1/R_2)^{1.5}$.  The distinction is included
here because the latter grouping produces a different quantity and can
reverse the sign for approximately flat rotation curves.

The same locked estimator can also be written, without changing its
computation, as
\begin{equation}
  \omega = \frac{V_2-V_{\rm Kep}(R_2)}{R_2}.
  \label{eq:omegaresidual}
\end{equation}
This algebraically equivalent form shows that $\omega$ is the outer-boundary
velocity residual relative to the Keplerian reference, normalized by $R_2$.
It is an interpretation of Eq.~\ref{eq:omega}, not a separately fitted
quantity.

\subsection{Units}

When $V$ is supplied in km~s$^{-1}$ and $R$ in kpc, the native output of
Eq.~\ref{eq:omega} is \omeganative.  All numerical $\omega$ values in
Tables~\ref{tab:upsilon} and \ref{tab:upsilon2} are reported in these
native computational units.  If a temporal-frequency representation is
desired, $1$~km~s$^{-1}$~kpc$^{-1}\simeq1.0227$~Gyr$^{-1}$.  Keeping
the native unit in the calculation makes Eq.~\ref{eq:vadj}
dimensionally transparent and prevents the approximate conversion
factor from being silently mixed into the velocity transformation.

\subsection{Baryonic velocity construction}

The baryonic velocity is constructed from SPARC component columns using
sign-preserving quadrature \cite{Corbelli2014}:
\begin{equation}
V_{\mathrm{bary}}(R)=
\sqrt{\mathrm{sgn}(V_g)V_g^2+\Upsilon V_d^2+\Upsilon V_b^2}.
\label{eq:vbary}
\end{equation}
The signed gas term preserves the convention used in mass-model tables
at inner radii where the tabulated gas contribution can be negative.

The maximum-disk upper bound used in our earlier analysis is
\begin{equation}
  \Upsilon_{\max}=\min_R\left[
  \frac{V_{\rm obs}^2-\mathrm{sgn}(V_g)V_g^2}
  {V_d^2+V_b^2+\varepsilon}\right]_{[0.1,1.0]},
  \label{eq:upsmax}
\end{equation}
where $\varepsilon=10^{-6}$~(km~s$^{-1}$)$^2$ prevents numerical
division by zero.  A secondary sensitivity analysis refines $\Upsilon$
within $[0.1,\Upsilon_{\max}]$ by bounded scalar minimization of
$\mathrm{RMSE}(V_{\rm adj},V_{\rm bary})$ using SciPy
\cite{scipy}.  Because that optimized $\Upsilon$ is selected using the
same residual later reported, it is treated as a sensitivity/upper-fit
analysis rather than an independent validation target.

\subsection{Pipeline}

For each galaxy the analysis performs the following deterministic
sequence:
\begin{enumerate}[leftmargin=2em]
\item load the ordered SPARC radial samples and baryonic component
columns;
\item select the first and last measured rotation-curve points;
\item compute $\omega$ with Eq.~\ref{eq:omega};
\item evaluate $V_{\rm Kep}(R)$ and $V_{\rm adj}(R)$ at every measured
radius;
\item construct $V_{\rm bary}(R)$ at $\Upsilon_{\max}$;
\item compute descriptive residual statistics;
\item optionally repeat the baryonic construction using bounded
$\Upsilon$ optimization as a sensitivity analysis;
\item generate the per-galaxy diagnostic panel and aggregate summary.
\end{enumerate}
No halo profile is fitted in this pipeline and no physical mechanism is
assigned to $\omega$.

\subsection{Numerical invariants and failure checks}
\label{sec:invariants}

Three identities provide useful regression tests for any implementation.
First, Eq.~\ref{eq:omega} must preserve its grouping exactly.  Second,
substitution at the outer boundary gives
\begin{equation}
  V_{\rm adj}(R_2)=V_1\sqrt{\frac{R_1}{R_2}}=V_{\rm Kep}(R_2).
  \label{eq:outeridentity}
\end{equation}
Thus the transformed outer endpoint is constrained by construction and
must not be presented as an independent validation datum.  Third,
$R\omega$ must return km~s$^{-1}$ when the native units above are used.
These checks are especially valuable because a parenthesis error in the
estimator changes both magnitude and sign while still producing
numerically plausible output.

\subsection{Uncertainty treatment}

With endpoint-velocity uncertainties $\sigma_{V_1}$ and
$\sigma_{V_2}$, and treating radii as fixed for this propagation, the
velocity contribution to the variance of $\omega$ is
\begin{equation}
\sigma_{\omega}^{2}=\left(\frac{\sigma_{V_2}}{R_2}\right)^2+
\left(\frac{\sigma_{V_1}}{R_1}\right)^2
\left(\frac{R_1}{R_2}\right)^3.
\label{eq:sigmaomega}
\end{equation}
For a general radial point, the corresponding variance is
\begin{equation}
\mathrm{Var}(V_{\rm adj})=\mathrm{Var}(V_{\rm obs})+R^2\mathrm{Var}(\omega)
-2R\,\mathrm{Cov}(V_{\rm obs},\omega).
\label{eq:varvadj}
\end{equation}
Away from the two endpoint measurements, the covariance term is neglected
under the same independence approximation.  At the endpoints it is non-zero
because $V_1$ and $V_2$ enter Eq.~\ref{eq:omega}; in particular,
$\mathrm{Cov}(V_2,\omega)=\sigma_{V_2}^2/R_2$.  Distance, inclination, and
stellar-population systematics are not fully propagated in the present
analysis.  In particular, optimizing $\Upsilon$ does not ``absorb''
those covariances in a statistical sense.  The reported RMSE values are
therefore descriptive reconstruction errors, not complete posterior
uncertainties.

\subsection{Contextual MOND calculation}

For context, our earlier analysis also evaluates the RAR form
\cite{McGaugh2016}
\begin{equation}
 g_{\rm obs}=\frac{g_{\rm bar}}{1-e^{-\sqrt{g_{\rm bar}/g_0}}},
 \qquad g_0=1.2\times10^{-10}\ {\rm m\,s^{-2}},
\end{equation}
with $\Upsilon=0.5$.  MOND and $\omega$ solve opposite mapping
problems: MOND predicts observed dynamics from baryons, whereas the
present transformation maps observed kinematics toward a baryonic
reference.  Consequently the MOND calculation is retained only as a
contextual benchmark, not as a headline model-selection test.

\section{Results}
\label{sec:results}

\subsection{Aggregate reconstruction error}

Table~\ref{tab:summary} reports the existing aggregate RMSE values.  Before
application of the transform, the mean observed--baryonic discrepancy is
51.82~\kms.  The primary maximum-disk arm uses no residual fit to
$V_{\rm bary}$: $\omega$ is fixed by the two observed boundary points and
$\Upsilon_{\max}$ is determined from the observed velocity and SPARC
component columns.  This zero-residual-fit transformation reduces the mean
$\mathrm{RMSE}(V_{\rm adj},V_{\rm bary})$ to 30.15~\kms.  Allowing the
bounded $\Upsilon$ sensitivity fit reduces the descriptive mean RMSE further
to 25.45~\kms.  The simple Keplerian reference used by the predecessor
analysis gives 74.20~\kms.  Because the optimized $\Upsilon$ is selected by
minimizing the same residual, the 25.45~\kms value should not be interpreted
as an out-of-sample prediction error.

\begin{table}[H]
\caption{Aggregate RMSE values from the 84-galaxy benchmark.  Rows with
different targets are retained for transparency but are not treated as
direct model-selection comparisons.\label{tab:summary}}
\centering
\small
\begin{tabular}{lccc}
\toprule
Calculation & Mean RMSE (\kms) & Target & Role\\
\midrule
Simple Kepler reference & 74.20 & $V_{\rm bary}$ & Reference\\
$\omega$, $\Upsilon_{\max}$ & 30.15 & $V_{\rm bary}$ & Primary reconstruction\\
$\omega$, $\Upsilon_{\rm opt}$ & 25.45 & $V_{\rm bary}$ & Sensitivity / fitted\\
MOND RAR & 18.19 & $V_{\rm obs}$ & MOND design target\\
MOND RAR & 60.57 & $V_{\rm bary}$ & Context only\\
Pure baryonic & 51.82 & $V_{\rm obs}$ & Standard baryonic gap\\
\bottomrule
\end{tabular}
\end{table}

\subsection{Per-galaxy reconstruction sensitivity}

\begin{figure}[H]
\centering
\includegraphics[width=0.96\textwidth]{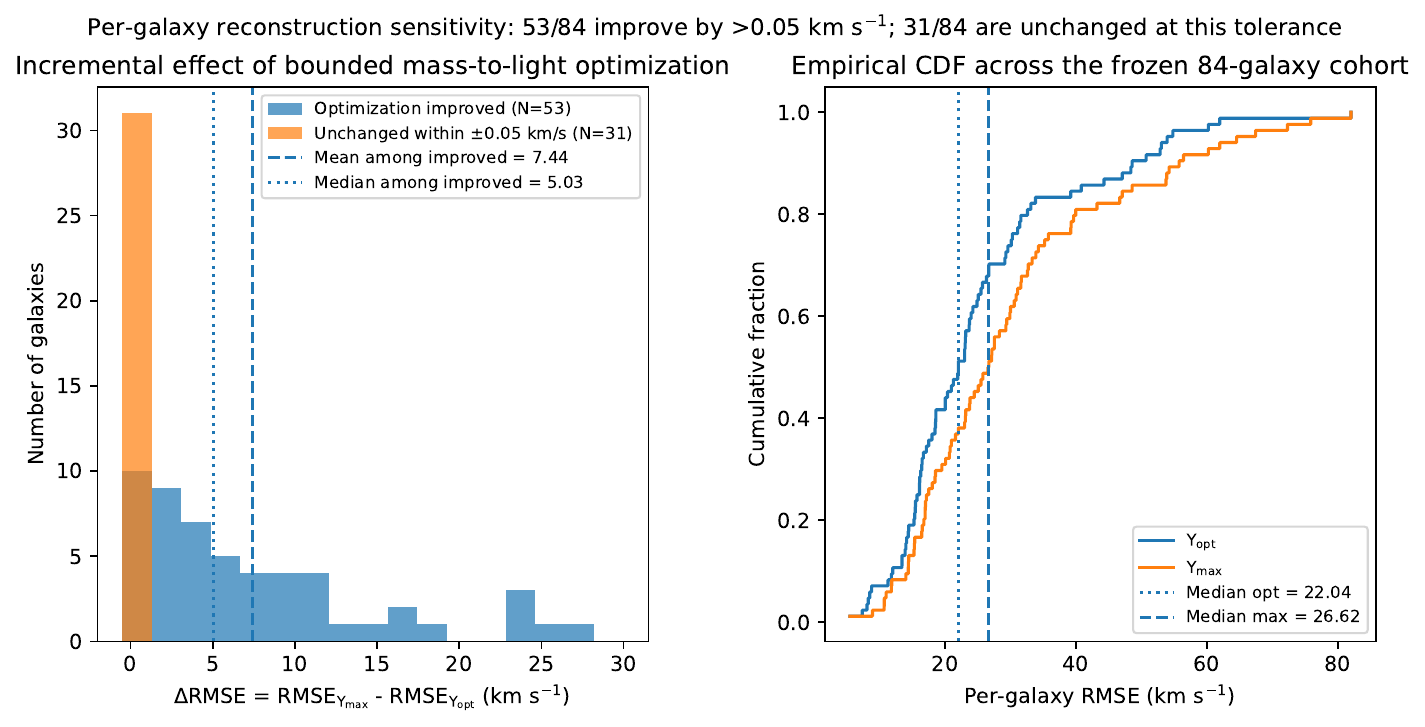}
\caption{Recomputed per-galaxy sensitivity statistics for the frozen
84-galaxy cohort.  Left: reduction in RMSE produced by bounded
$\Upsilon$ optimization; improvements are counted only when
$\Delta\mathrm{RMSE}>0.05$~\kms, yielding 53 improved and 31 unchanged
systems at the tabulated precision.  Right: empirical CDFs of the
per-galaxy RMSE values for the primary $\Upsilon_{\max}$ reconstruction and
the fitted $\Upsilon_{\rm opt}$ sensitivity arm.}
\label{fig:improvementcdf}
\end{figure}

The aggregate means conceal substantial galaxy-to-galaxy variation, so we
reconstructed the per-galaxy sensitivity statistics directly from the 84 rows
of Tables~\ref{tab:upsilon} and \ref{tab:upsilon2}.  To prevent differences
at the displayed rounding precision from being counted as numerical
improvements, we classify an optimization benefit only when
$\Delta\mathrm{RMSE}=\mathrm{RMSE}_{\Upsilon_{\max}}-
\mathrm{RMSE}_{\Upsilon_{\rm opt}}>0.05$~\kms.  Under that explicit rule,
53 of 84 galaxies show an optimization benefit and 31 are unchanged at the
tabulated precision.  NGC~3198 changes from 47.07 to 47.08~\kms; this
$-0.01$~\kms difference is therefore classified as unresolved rather than
as a regression under the stated tolerance.  The $\Upsilon$ columns are
displayed to three decimals and the full-precision optimizer state is not
preserved in the published table, so equality of the displayed values does
not establish equality of the underlying full-precision optimizer inputs.
Across all 84 systems the mean RMSE changes from 30.148 to 25.455~\kms and
the median changes from 26.62 to 22.04~\kms.  Among the 53 galaxies with a
resolved optimization benefit, the mean and median reductions are 7.44 and
5.03~\kms, respectively.  Figure~\ref{fig:improvementcdf} shows both the
improvement distribution and the empirical CDF of the two reconstruction
arms.  These values are reconstructed from the manuscript's frozen
84-galaxy numerical table rather than inherited from a legacy plotting
script.

\subsection{Full-sample visual diagnostics}

\begin{figure}[H]
\centering
\includegraphics[width=0.9\textwidth]{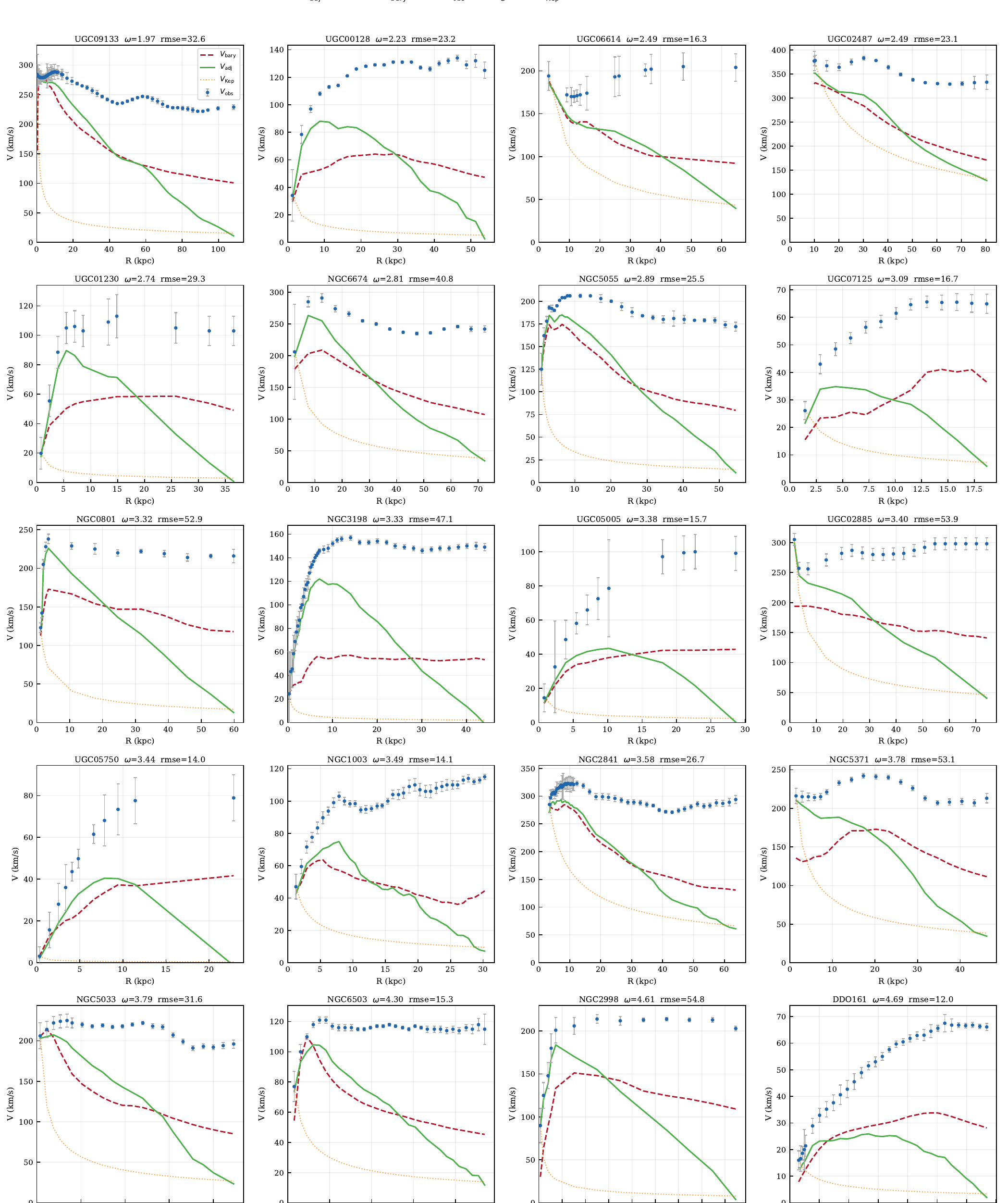}
\caption{First 20 galaxies in the frozen 84-galaxy benchmark.  Blue:
$V_{\rm obs}$; green: $V_{\rm adj}$; orange-dotted: $V_{\rm Kep}$;
red-dashed: optimized $V_{\rm bary}$.  The figure is retained from the
source analysis as a direct diagnostic of the transformation.}
\label{fig:first20}
\end{figure}

\begin{figure}[H]
\centering
\includegraphics[width=\textwidth,height=0.88\textheight,keepaspectratio]{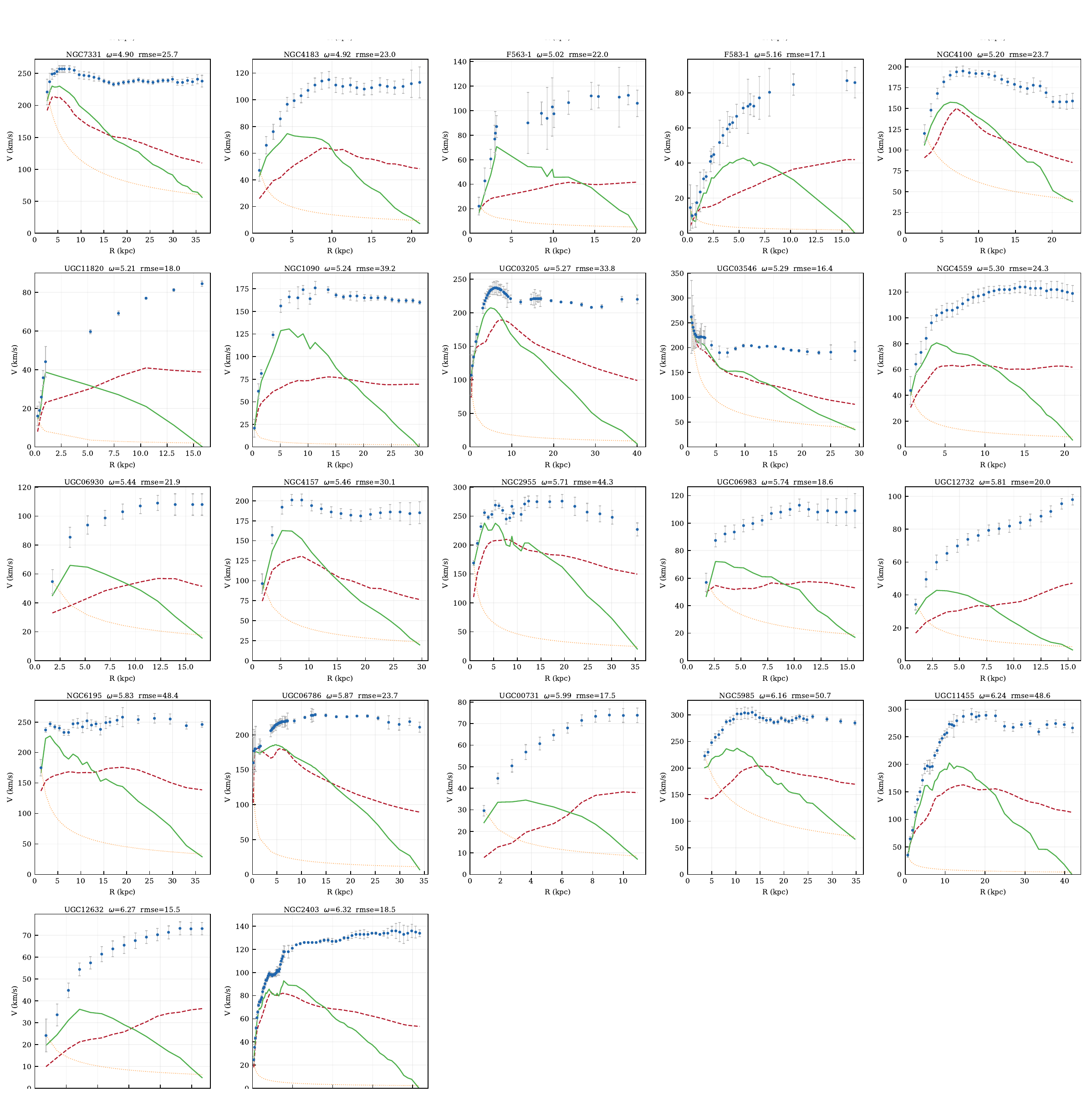}
\caption{Galaxies 21--42 of the 84-galaxy benchmark, sorted by the
reported native $\omega$ value.  Colour scheme as in
Fig.~\ref{fig:first20}.}
\label{fig:all84a}
\end{figure}

\begin{figure}[H]
\centering
\includegraphics[width=\textwidth,height=0.88\textheight,keepaspectratio]{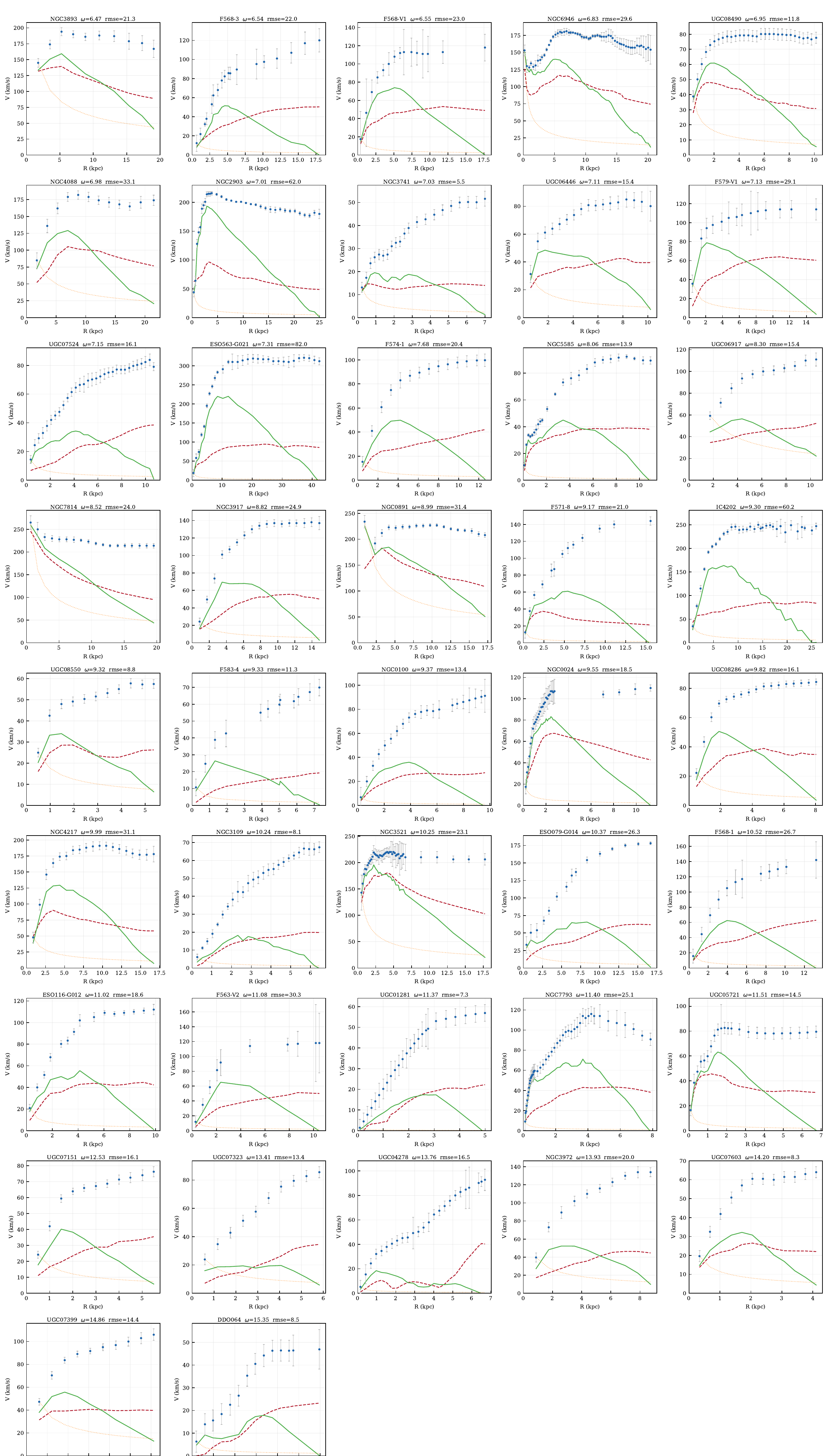}
\caption{Galaxies 43--84 of the 84-galaxy benchmark, sorted by the
reported native $\omega$ value.  Colour scheme as in
Fig.~\ref{fig:first20}.}
\label{fig:all84b}
\end{figure}

Figures~\ref{fig:first20}--\ref{fig:all84b} retain the complete visual
record of the benchmark.  Every panel shows the observed curve, the
transformed curve, the Keplerian reference, and the baryonic
reconstruction.  These figures are intentionally preserved rather than
reduced to a small set of favourable examples: the computational claim
is a sample-wide transformation whose failures and mismatches must
remain visible.

\subsection{Log-slope and enclosed-mass diagnostics}

\begin{figure}[H]
\centering
\includegraphics[width=0.9\textwidth]{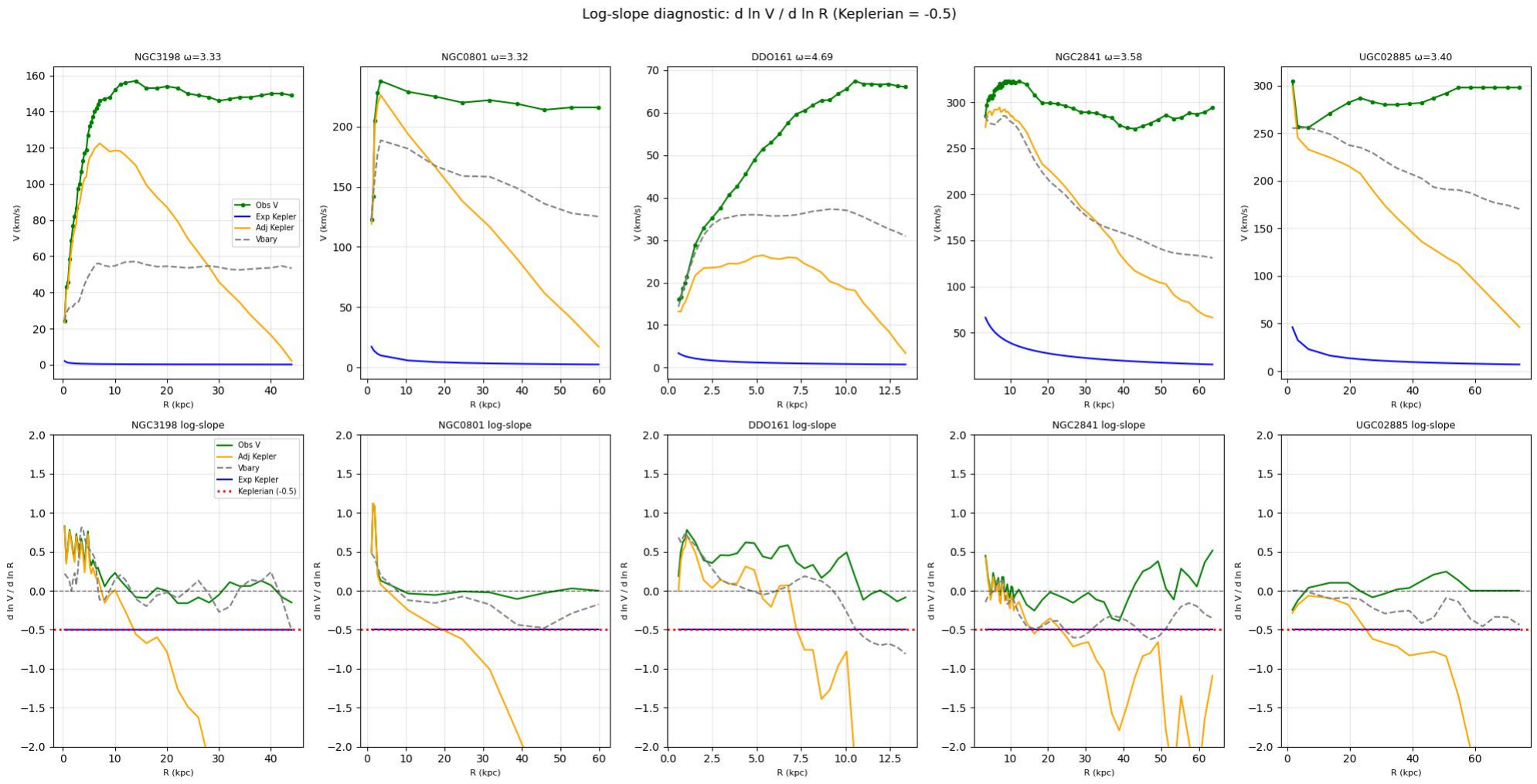}
\caption{Log-slope $\DD\ln V/\DD\ln R$ for five representative
systems.  The red dashed line marks $-1/2$.  The strong decline in the
transformed curve is expected from the algebra of Eq.~\ref{eq:vadj} and
is used here as a diagnostic rather than an independent physical test.}
\label{fig:logslope}
\end{figure}

The log-slope plots in Fig.~\ref{fig:logslope} show how the linear
$R\omega$ term reshapes the radial derivative.  For a flat observed
curve $V_{\rm obs}=V_0$,
\begin{equation}
\frac{\DD\ln V_{\rm adj}}{\DD\ln R}
=-\frac{R\omega}{V_0-R\omega},
\end{equation}
so the slope becomes rapidly negative as the transformed velocity
approaches zero.  This behaviour is a mathematical property of the
transformation and should not by itself be interpreted as evidence for
a physical Keplerian regime.

\begin{figure}[H]
\centering
\includegraphics[width=0.9\textwidth]{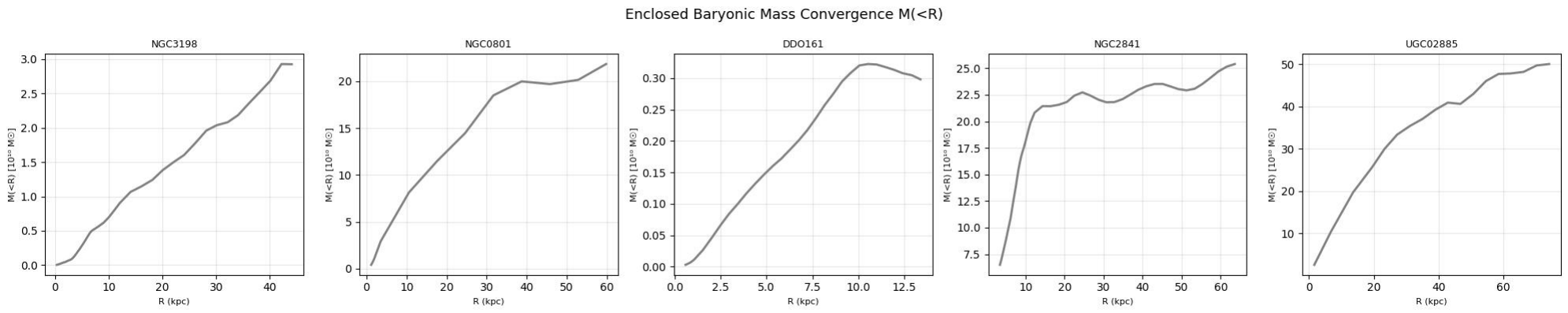}
\caption{Enclosed baryonic mass $M(<R)$ for five representative
galaxies.  UGC~02885 remains rising at the outer measured radius and is
therefore a useful stress case for the two-boundary approximation.}
\label{fig:enclosed}
\end{figure}

Figure~\ref{fig:enclosed} shows the enclosed baryonic-mass diagnostic
computed as $M(<R)=V_{\rm bary}^2R/G$.  Most displayed cases flatten or
begin to converge over the measured radial interval; UGC~02885 remains
an important counterexample with rising enclosed mass at the last
measured point.  This diagnostic identifies where an outer Keplerian
reference is least defensible.

\subsection{Boundary behaviour}

Our earlier analysis finds $V_{\rm adj}(R_2)<V_{\rm bary}(R_2)$ for
all 84 benchmark galaxies.  We retain those values in
Table~\ref{tab:outergap} as a descriptive boundary diagnostic, but do
not treat the sign as independent validation: Eq.~\ref{eq:outeridentity}
forces $V_{\rm adj}(R_2)=V_{\rm Kep}(R_2)$ by construction.  The table
therefore characterizes the relationship between the imposed outer
reference and the baryonic model, rather than demonstrating the absence
or presence of a halo component.

\section{Discussion}
\label{sec:discussion}

\subsection{What is validated}

The strongest result of the present work is narrower than the language
of the earlier analysis.  The 84-panel analysis demonstrates that the
canonical two-boundary transform is computationally stable enough to
produce a coherent family of radial reconstructions across a diverse
SPARC benchmark, and that those reconstructions are substantially
closer to the assembled baryonic curves than the simple Keplerian
reference in the reported RMSE summary.  That is a validation of a
\emph{data transformation and diagnostic workflow}.  It is not a
validation of a new force law, a replacement for dark matter, or a
unique physical interpretation of $\omega$.

This distinction also protects the principal scientific mission of
\cite{Flynn2025}.  The predecessor paper asked whether a compact
kinematic correction could reproduce the characteristic rotation-curve
morphology exemplified by M33.  The present paper tests the same
correction against resolved baryonic component information rather than
changing the estimator to obtain a better result.

\subsection{Mass-to-light-ratio optimization}

The bounded $\Upsilon$ fit improves the descriptive RMSE from 30.15 to
25.45~\kms, but it cannot serve as an independent validation because it
uses $V_{\rm adj}$ in its objective function.  We therefore make the
maximum-disk result the primary reconstruction and retain the optimized
result as a sensitivity analysis.  The fitted values also show material
boundary behavior: 8 of 84 solutions terminate at the imposed lower bound
$\Upsilon_{\rm opt}=0.1$, the median $\Upsilon_{\rm opt}$ is 0.428, and
28 of 84 fitted values are below 0.3.  For comparison, the median
$\Upsilon_{\max}$ is 0.835 and 29 of 84 systems reach the imposed upper
bound of 1.0.  These features reinforce the interpretation of the optimized
arm as a bounded sensitivity fit rather than an independent physical
mass-to-light determination.  This separation is preferable to claiming that
$\Upsilon$ optimization absorbs distance, inclination, or stellar-population
uncertainty; those systematics require explicit propagation or Monte Carlo
treatment in a future analysis.

\subsection{MOND and halo models}

The current data product does not support a symmetric model-selection
contest among $\omega$, MOND, and $\Lambda$CDM.  MOND maps baryonic
acceleration to the observed field, while the $\omega$ procedure studied
here maps observed kinematics toward a baryonic reference.  Similarly,
a fair NFW comparison would require per-galaxy halo fits with explicit
parameter counts and an agreed comparison target.  We therefore retain
the MOND numbers only to document our earlier analysis and defer formal
model selection to work designed around equal targets and degrees of
freedom.

\subsection{Failure modes and applicability}

The method is most vulnerable when either boundary point is poorly
sampled, the outer radius has not reached approximate enclosed-mass
convergence, or strong non-circular motions invalidate the use of a
single ordered rotation curve.  UGC~02885 illustrates the second case.
The source analysis identifies six galaxies for which the transformed
curve does not beat the simple Keplerian reference: ESO563-G021,
NGC~4217, IC~4202, NGC~2903, NGC~7793, and F571-8.  We reconstructed the
context of those six cases from the same 84-row numerical table used for
Fig.~\ref{fig:improvementcdf}.  All six have
$\Upsilon_{\max}\leq0.111$, with four at the imposed 0.1 floor, whereas
their $\omega$ values span the interior of the sample distribution rather
than its high-$\omega$ tail.  This is a post-hoc association, not a
selection rule or causal test, but it localizes the observed failures near
a regime in which the baryonic-disk upper bound is itself highly
restrictive.

\begin{figure}[H]
\centering
\includegraphics[width=0.96\textwidth]{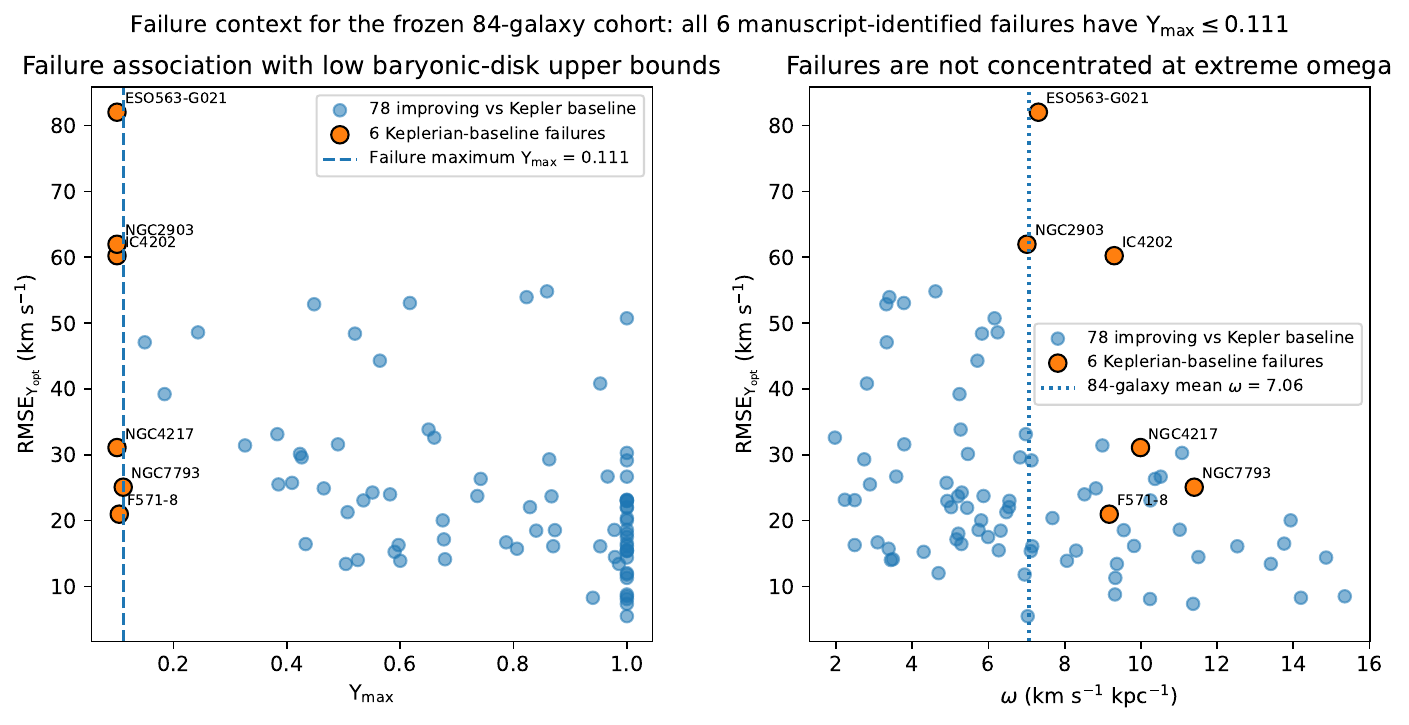}
\caption{Recomputed context for the six Keplerian-reference non-improvements
identified in the source analysis.  Left: all six lie at
$\Upsilon_{\max}\leq0.111$ (four at the 0.1 floor).  Right: the same six
systems are not concentrated at extreme $\omega$.  The association with
low $\Upsilon_{\max}$ is descriptive and post hoc; it is not used to
exclude galaxies or tune the transformation.}
\label{fig:failurecontext}
\end{figure}

The frozen 84-galaxy cohort is retained unchanged throughout this
submission to preserve direct continuity with the predecessor validation.
No galaxies are added to or removed from the benchmark during the present
reconstruction.

\subsection{Reproducibility and software relevance}

The practical value of this analysis for astronomical computing is that
a small algebraic expression sits at the root of every downstream
figure and statistic.  A grouping error can therefore contaminate an
entire multi-survey interpretation while leaving superficially
reasonable outputs.  Explicit computational units, cohort membership,
endpoint identities, and formula-level regression tests are not merely
software-engineering conveniences; in this setting they are part of the
scientific method.  The deposited scripts and derived products make the
pipeline independently inspectable, while the figures preserve the
full-sample visual evidence rather than only aggregate metrics.

\section{Conclusions}
\label{sec:conclusions}

The baryonic follow-up to \cite{Flynn2025} can be stated conservatively
and reproducibly:
\begin{enumerate}[leftmargin=2em]
\item The canonical estimator is
$\omega=V_2/R_2-(V_1/R_1)(R_1/R_2)^{3/2}$ and must not be replaced by a
mis-parenthesized difference-of-ratios expression.
\item In native units (km~s$^{-1}$~kpc$^{-1}$), the transform
$V_{\rm adj}=V_{\rm obs}-R\omega$ is dimensionally direct.
\item Across the frozen 84-galaxy benchmark, the maximum-disk
reconstruction has a mean $V_{\rm adj}$--$V_{\rm bary}$ RMSE of
30.15~\kms; bounded $\Upsilon$ optimization reduces the descriptive
value to 25.45~\kms.  Recalculation from the tabulated 84-galaxy results
shows a resolved mass-to-light optimization benefit ($\Delta\mathrm{RMSE}>0.05$~\kms)
in 53 galaxies and no resolved change in 31, compared with 74.20~\kms
for the simple Keplerian
reference in our earlier analysis.
\item The optimized $\Upsilon$ result is a sensitivity fit, not an
independent baryonic validation target.
\item The outer endpoint and the steep transformed log-slope have
algebraic dependencies on the estimator and are therefore diagnostics,
not independent physical evidence.
\item The complete 84-galaxy panel set is retained so that both successes
and failures remain visible and reproducible.
\end{enumerate}

These results support continued use of $\omega$ as an empirical
kinematic diagnostic whose physical interpretation remains open.  The
next scientifically clean extensions are to propagate distance and
inclination systematics and to apply the locked estimator, without
algebraic modification, to independent resolved rotation-curve data
sets.

\section*{CRediT authorship contribution statement}
\textbf{D.\,C. Flynn:} Conceptualization, Data curation, Formal
analysis, Investigation, Methodology, Software, Validation,
Visualization, Writing -- original draft, Writing -- review and editing.

\section*{Declaration of competing interests}
The author declares no competing interests.

\section*{Funding}
This research did not receive any specific grant from funding agencies
in the public, commercial, or not-for-profit sectors.

\section*{Data and code availability}
All observational inputs are drawn from the publicly available SPARC
database \cite{Lelli2016}.  The computation scripts and derived data
products associated with the 84-galaxy analysis are deposited
at Zenodo (\url{https://doi.org/10.5281/zenodo.19798527}) under CC~BY~4.0.
For reproducibility, the exact benchmark galaxy list is also contained
in Tables~\ref{tab:upsilon} and \ref{tab:upsilon2} of this manuscript.

\section*{Acknowledgements}
The SPARC database is maintained by F. Lelli and S. McGaugh (CWRU).
This work was conducted as independent research by EPS Research and
received no external funding.

\section*{Declaration of generative AI and AI-assisted technologies in the manuscript preparation process}
During preparation of the manuscript, generative AI systems were used
for manuscript review, LaTeX formatting, literature cross-checking, and
language editing.  The author reviewed and edited the resulting text and
takes full responsibility for the content.  Numerical results, tables,
and figures in the source analysis were computed in JupyterLab from
SPARC source files; the reproducible analysis products are deposited at
the Zenodo record cited above.  No AI system is listed as an author.

\appendix
\renewcommand{\thetable}{A.\arabic{table}}
\setcounter{table}{1}
\begin{table}[!p]
\caption{Per-galaxy $\Upsilon$ optimisation, 84 SPARC $Q\!=\!1$
galaxies, sorted by RMSE improvement. $\omega$ in
km~s$^{-1}$~kpc$^{-1}$; RMSE in \kms.\label{tab:upsilon}}
\centering
\footnotesize
\renewcommand{\arraystretch}{0.94}
\begin{tabular}{lrrrrrl}
\toprule
Galaxy & $\omega$ & $\Upsilon_{\max}$ & $\Upsilon_{\mathrm{opt}}$ &
RMSE$_{\mathrm{old}}$ & RMSE$_{\mathrm{new}}$ & $\Delta$\\
\midrule
ESO079-G014  & 10.37 & 0.742 & 0.142 & 53.87 & 26.33 & $+27.5$\\
UGC07323     & 13.41 & 0.986 & 0.100 & 39.27 & 13.42 & $+25.9$\\
NGC4088      &  6.98 & 0.383 & 0.145 & 56.43 & 33.12 & $+23.3$\\
NGC3972      & 13.93 & 0.675 & 0.127 & 43.21 & 20.03 & $+23.2$\\
NGC5371      &  3.78 & 0.617 & 0.317 & 75.88 & 53.05 & $+22.8$\\
UGC02885     &  3.40 & 0.823 & 0.474 & 72.32 & 53.94 & $+18.4$\\
NGC6195      &  5.83 & 0.520 & 0.318 & 64.55 & 48.40 & $+16.1$\\
NGC4100      &  5.20 & 0.867 & 0.516 & 39.64 & 23.68 & $+16.0$\\
F583-4       &  9.33 & 1.000 & 0.100 & 25.77 & 11.29 & $+14.5$\\
NGC2998      &  4.61 & 0.859 & 0.460 & 67.46 & 54.82 & $+12.6$\\
UGC04278     & 13.76 & 1.000 & 0.111 & 28.30 & 16.50 & $+11.8$\\
UGC06917     &  8.30 & 1.000 & 0.379 & 27.12 & 15.44 & $+11.7$\\
UGC07151     & 12.53 & 0.953 & 0.200 & 27.53 & 16.09 & $+11.4$\\
ESO116-G012  & 11.02 & 1.000 & 0.304 & 29.42 & 18.61 & $+10.8$\\
NGC3521      & 10.25 & 0.535 & 0.404 & 33.30 & 23.06 & $+10.2$\\
NGC4157      &  5.46 & 0.423 & 0.253 & 39.98 & 30.12 & $+9.9$\\
NGC2955      &  5.71 & 0.564 & 0.411 & 53.77 & 44.29 & $+9.5$\\
NGC7331      &  4.90 & 0.409 & 0.310 & 34.27 & 25.73 & $+8.5$\\
UGC06930     &  5.44 & 1.000 & 0.394 & 29.91 & 21.94 & $+8.0$\\
NGC3917      &  8.82 & 0.465 & 0.192 & 32.71 & 24.89 & $+7.8$\\
NGC4183      &  4.92 & 1.000 & 0.437 & 30.00 & 22.98 & $+7.0$\\
F574-1       &  7.68 & 1.000 & 0.344 & 27.16 & 20.38 & $+6.8$\\
NGC4559      &  5.30 & 0.551 & 0.270 & 30.57 & 24.26 & $+6.3$\\
NGC6674      &  2.81 & 0.953 & 0.719 & 46.70 & 40.82 & $+5.9$\\
NGC6946      &  6.83 & 0.426 & 0.289 & 35.22 & 29.60 & $+5.6$\\
NGC0100      &  9.37 & 0.504 & 0.170 & 18.46 & 13.41 & $+5.1$\\
NGC5985      &  6.16 & 1.000 & 0.786 & 55.76 & 50.73 & $+5.0$\\
UGC07524     &  7.15 & 0.870 & 0.195 & 20.78 & 16.10 & $+4.7$\\
NGC0891      &  8.99 & 0.326 & 0.257 & 35.78 & 31.41 & $+4.4$\\
UGC07125     &  3.09 & 0.787 & 0.208 & 20.09 & 16.69 & $+3.4$\\
UGC01281     & 11.37 & 1.000 & 0.228 & 10.72 & 7.35 & $+3.4$\\
DDO064       & 15.35 & 1.000 & 0.163 & 11.86 & 8.49 & $+3.4$\\
DDO161       &  4.69 & 1.000 & 0.100 & 15.32 & 12.01 & $+3.3$\\
NGC2403      &  6.32 & 0.840 & 0.597 & 21.59 & 18.47 & $+3.1$\\
UGC07603     & 14.20 & 0.940 & 0.437 & 11.00 & 8.27 & $+2.7$\\
NGC5585      &  8.06 & 0.600 & 0.319 & 16.52 & 13.89 & $+2.6$\\
NGC3109      & 10.24 & 1.000 & 0.100 & 10.69 & 8.08 & $+2.6$\\
NGC7814      &  8.52 & 0.582 & 0.501 & 26.55 & 23.98 & $+2.6$\\
NGC3893      &  6.47 & 0.507 & 0.417 & 23.82 & 21.27 & $+2.6$\\
F579-V1      &  7.13 & 1.000 & 0.633 & 31.66 & 29.14 & $+2.5$\\
F568-3       &  6.54 & 0.829 & 0.423 & 24.46 & 22.04 & $+2.4$\\
UGC12632     &  6.27 & 1.000 & 0.476 & 17.02 & 15.48 & $+1.5$\\
\bottomrule
\end{tabular}
\end{table}

\begin{table}[!p]
\caption{Per-galaxy $\Upsilon$ optimisation (continued).\label{tab:upsilon2}}
\centering
\footnotesize
\renewcommand{\arraystretch}{0.94}
\begin{tabular}{lrrrrrl}
\toprule
Galaxy & $\omega$ & $\Upsilon_{\max}$ & $\Upsilon_{\mathrm{opt}}$ &
RMSE$_{\mathrm{old}}$ & RMSE$_{\mathrm{new}}$ & $\Delta$\\
\midrule
NGC0801      &  3.32 & 0.448 & 0.377 & 54.24 & 52.86 & $+1.4$\\
UGC05005     &  3.38 & 0.806 & 0.475 & 16.97 & 15.72 & $+1.2$\\
UGC06983     &  5.74 & 0.978 & 0.749 & 19.52 & 18.58 & $+0.9$\\
F568-1       & 10.52 & 1.000 & 0.684 & 27.53 & 26.67 & $+0.9$\\
UGC08286     &  9.82 & 1.000 & 0.690 & 16.95 & 16.14 & $+0.8$\\
UGC12732     &  5.81 & 1.000 & 0.656 & 20.66 & 20.03 & $+0.6$\\
F563-V2      & 11.08 & 1.000 & 0.661 & 30.87 & 30.28 & $+0.6$\\
UGC06614     &  2.49 & 0.597 & 0.561 & 16.76 & 16.28 & $+0.5$\\
NGC1003      &  3.49 & 0.679 & 0.583 & 14.38 & 14.12 & $+0.3$\\
UGC08550     &  9.32 & 1.000 & 0.848 & 8.91 & 8.77 & $+0.1$\\
UGC03205     &  5.27 & 0.650 & 0.634 & 33.88 & 33.83 & $+0.1$\\
UGC11820     &  5.21 & 1.000 & 0.871 & 18.05 & 18.01 & $0.0$\\
UGC07399     & 14.86 & 1.000 & 0.963 & 14.39 & 14.38 & $0.0$\\
ESO563-G021  &  7.31 & 0.100 & 0.100 & 82.04 & 82.04 & $0.0$\\
NGC4217      &  9.99 & 0.100 & 0.100 & 31.09 & 31.09 & $0.0$\\
IC4202       &  9.30 & 0.100 & 0.100 & 60.24 & 60.24 & $0.0$\\
NGC2903      &  7.01 & 0.100 & 0.100 & 61.98 & 61.98 & $0.0$\\
UGC00731     &  5.99 & 1.000 & 1.000 & 17.50 & 17.50 & $0.0$\\
NGC3741      &  7.03 & 1.000 & 1.000 & 5.47 & 5.47 & $0.0$\\
NGC6503      &  4.30 & 0.590 & 0.589 & 15.25 & 15.25 & $0.0$\\
UGC06446     &  7.11 & 1.000 & 1.000 & 15.36 & 15.36 & $0.0$\\
NGC2841      &  3.58 & 0.966 & 0.963 & 26.69 & 26.69 & $0.0$\\
UGC05750     &  3.44 & 0.525 & 0.525 & 14.01 & 14.01 & $0.0$\\
UGC08490     &  6.95 & 1.000 & 1.000 & 11.80 & 11.80 & $0.0$\\
F563-1       &  5.02 & 1.000 & 1.000 & 22.03 & 22.03 & $0.0$\\
UGC05721     & 11.51 & 0.979 & 0.979 & 14.46 & 14.46 & $0.0$\\
UGC00128     &  2.23 & 1.000 & 1.000 & 23.16 & 23.16 & $0.0$\\
UGC01230     &  2.74 & 0.863 & 0.863 & 29.30 & 29.30 & $0.0$\\
F583-1       &  5.16 & 0.677 & 0.677 & 17.14 & 17.14 & $0.0$\\
F568-V1      &  6.55 & 1.000 & 1.000 & 23.00 & 23.00 & $0.0$\\
NGC0024      &  9.55 & 0.873 & 0.873 & 18.54 & 18.54 & $0.0$\\
NGC5055      &  2.89 & 0.385 & 0.385 & 25.49 & 25.49 & $0.0$\\
UGC06786     &  5.87 & 0.736 & 0.736 & 23.73 & 23.73 & $0.0$\\
UGC03546     &  5.29 & 0.433 & 0.433 & 16.43 & 16.43 & $0.0$\\
NGC5033      &  3.79 & 0.490 & 0.490 & 31.58 & 31.58 & $0.0$\\
UGC02487     &  2.49 & 1.000 & 1.000 & 23.10 & 23.10 & $0.0$\\
UGC09133     &  1.97 & 0.660 & 0.660 & 32.60 & 32.60 & $0.0$\\
NGC1090      &  5.24 & 0.184 & 0.184 & 39.21 & 39.21 & $0.0$\\
UGC11455     &  6.24 & 0.243 & 0.243 & 48.59 & 48.59 & $0.0$\\
NGC7793      & 11.40 & 0.111 & 0.111 & 25.07 & 25.07 & $0.0$\\
F571-8       &  9.17 & 0.104 & 0.104 & 20.97 & 20.97 & $0.0$\\
NGC3198      &  3.33 & 0.149 & 0.149 & 47.07 & 47.08 & $-0.01$\\
\bottomrule
\end{tabular}
\end{table}
\FloatBarrier
\renewcommand{\thetable}{B.\arabic{table}}
\setcounter{table}{3}
\section{Derived boundary and contextual comparison tables}

\begin{table}[H]
\caption{Outer gap $V_{\mathrm{adj}}(R_2) - V_{\mathrm{bary}}(R_2)$,
selected galaxies. This is a derived boundary diagnostic, not an independent validation test.
All 84 gaps negative. Mean $-51.4 \pm 25.0$~\kms.\label{tab:outergap}}
\centering
\small
\begin{tabular}{lrrr}
\toprule
Galaxy & $V_{\mathrm{adj}}$~(\kms) & $V_{\mathrm{bary}}$~(\kms) &
Gap~(\kms)\\
\midrule
UGC09133 &  18.0 & 107.8 & $-89.8$\\
NGC0801  &  17.9 & 122.8 & $-104.9$\\
NGC2955  &  24.3 & 153.8 & $-129.5$\\
UGC02885 &  45.7 & 146.1 & $-100.4$\\
NGC3198  &   2.0 &  56.4 & $-54.4$\\
DDO161   &   3.2 &  29.3 & $-26.1$\\
NGC2841  &  65.1 & 134.8 & $-69.7$\\
NGC3741  &   2.3 &  15.0 & $-12.7$\\
\midrule
\multicolumn{4}{l}{All 84 negative; mean $-51.4$, SD $25.0$~\kms}\\
\bottomrule
\end{tabular}
\end{table}

\begin{table}[H]
\caption{Four-way RMSE comparison, selected galaxies (full table at
CDS). $\omega$ vs \Vbary; MOND vs \Vobs\ (design target) and vs
\Vbary\ (same-target comparison); baryonic vs \Vobs. Rows with unlike scientific targets are retained for context and are not used here as a formal model-selection test.\label{tab:mond}}
\centering
\small
\begin{tabular}{lrrrrr}
\toprule
Galaxy & $\omega$ & RMSE$_\omega$ & RMSE$_{\mathrm{M,Vo}}$ &
  RMSE$_{\mathrm{M,Vb}}$ & RMSE$_{\mathrm{b}}$\\
 & (km~s$^{-1}$~kpc$^{-1}$) & vs \Vbary & vs \Vobs & vs \Vbary & vs \Vobs\\
\midrule
UGC06614 & 2.5 & 16.3 & 16.6 &  78.4 &  66.2\\
UGC02487 & 2.5 & 23.1 & 71.6 &  64.1 & 109.6\\
NGC5055  & 2.9 & 25.5 & 27.6 &  89.9 &  64.1\\
DDO161   & 4.7 & 12.0 & 18.5 &  43.2 &  21.3\\
NGC2841  & 3.6 & 26.7 & 64.6 &  59.9 &  90.3\\
NGC3198  & 3.3 & 47.1 & 10.9 &  88.3 &  82.1\\
\midrule
\multicolumn{6}{l}{Mean: $\omega$ 25.45 (vs \Vbary); MOND 18.19
  (vs \Vobs); MOND 60.57 (vs \Vbary); bary 51.82~\kms\ (vs \Vobs)}\\
\bottomrule
\end{tabular}
\end{table}

% Bibliography embedded directly for a self-contained submission source.


\begin{thebibliography}{10}
\expandafter\ifx\csname url\endcsname\relax
  \def\url#1{\texttt{#1}}\fi
\expandafter\ifx\csname urlprefix\endcsname\relax\def\urlprefix{URL }\fi
\expandafter\ifx\csname href\endcsname\relax
  \def\href#1#2{#2} \def\path#1{#1}\fi

\bibitem{Rubin1980}
V.~C. Rubin, W.~K. Ford, N.~Thonnard, Rotational properties of 21 {SC} galaxies
  with a large range of luminosities and radii, ApJ 238 (1980) 471.

\bibitem{vanAlbada1985}
T.~S. {van Albada}, J.~N. Bahcall, K.~Begeman, R.~Sancisi, Distribution of dark
  matter in the spiral galaxy {NGC} 3198, ApJ 295 (1985) 305.

\bibitem{NFW1997}
J.~F. Navarro, C.~S. Frenk, S.~D.~M. White, A universal density profile from
  hierarchical clustering, ApJ 490 (1997) 493.

\bibitem{Begeman1991}
K.~G. Begeman, A.~H. Broeils, R.~H. Sanders, Extended rotation curves of spiral
  galaxies: dark haloes and modified dynamics, MNRAS 249 (1991) 523.

\bibitem{Milgrom1983}
M.~Milgrom, A modification of the {Newtonian} dynamics as a possible
  alternative to the hidden mass hypothesis, ApJ 270 (1983) 365.

\bibitem{Sanders2002}
R.~H. Sanders, S.~S. McGaugh, Modified {Newtonian} dynamics as an alternative
  to dark matter, ARA\&A 40 (2002) 263.

\bibitem{FamaeySanders2012}
B.~Famaey, S.~S. McGaugh, Modified {Newtonian} dynamics ({MOND}): Observational
  phenomenology and relativistic extensions, Living Reviews in Relativity 15
  (2012) 10.

\bibitem{McGaugh2016}
S.~S. McGaugh, F.~Lelli, J.~M. Schombert, Radial acceleration relation in
  rotationally supported galaxies, Phys.\ Rev.\ Lett. 117 (2016) 201101.

\bibitem{Lelli2016}
F.~Lelli, S.~S. McGaugh, J.~M. Schombert, {SPARC}: Mass models for 175 disk
  galaxies with {Spitzer} photometry and accurate rotation curves, AJ 152
  (2016) 157.

\bibitem{Lelli2017}
F.~Lelli, S.~S. McGaugh, J.~M. Schombert, M.~S. Pawlowski, One law to rule them
  all: The radial acceleration relation of galaxies, ApJ 836 (2017) 152.

\bibitem{Walter2008}
F.~Walter, E.~Brinks, W.~J.~G. {de Blok}, et~al., {THINGS}: The {H\,\textsc{i}}
  nearby galaxy survey, AJ 136 (2008) 2563.

\bibitem{deBlok2008}
W.~J.~G. {de Blok}, F.~Walter, E.~Brinks, et~al., {THINGS}: The {H\,\textsc{i}}
  nearby galaxy survey, AJ 136 (2008) 2648.

\bibitem{Oh2015}
S.-H. Oh, D.~A. Hunter, E.~Brinks, et~al., {LITTLE THINGS} in {3D}: Robust
  determination of the circular velocity of dwarf irregular galaxies, AJ 149
  (2015) 180.

\bibitem{Corbelli2014}
E.~Corbelli, D.~Thilker, S.~Zibetti, C.~Giovanardi, P.~Salucci, Growing a
  bulgeless galaxy from cold streams, A\&A 572 (2014) A23.

\bibitem{Flynn2025}
D.~C. Flynn, J.~Cannaliato, A new empirical fit to galaxy rotation curves,
  Frontiers in Astronomy and Space Sciences 12 (2025).
\newblock \href {https://doi.org/10.3389/fspas.2025.1680387}
  {\path{doi:10.3389/fspas.2025.1680387}}.

\bibitem{scipy}
P.~Virtanen, R.~Gommers, T.~E. Oliphant, et~al., {SciPy} 1.0: fundamental
  algorithms for scientific computing in {Python}, Nature Methods 17 (2020)
  261.

\end{thebibliography}
\end{document}